\documentclass[fleqn,twoside]{article}
\usepackage{espcrc2}
\usepackage{amssymb}
\usepackage{graphicx}
\usepackage{latexsym}
\usepackage[figuresright]{rotating}
\newcommand{\eq}{\begin{equation}}
\newcommand{\en}{\end{equation}}
\newcommand{\qe}{\end{equation}}
\newcommand{\ear}{\begin{eqnarray}}
\newcommand{\eqa}{\begin{eqnarray}}
\newcommand{\rae}{\end{eqnarray}}
\newcommand{\ena}{\end{eqnarray}}
\newcommand{\beq}{\begin{equation}}
\newcommand{\eeq}{\end{equation}}
\newcommand{\bea}{\begin{eqnarray}}
\newcommand{\eea}{\end{eqnarray}}

\newcommand{\AmS}{{\protect\the\textfont2
  A\kern-.1667em\lower.5ex\hbox{M}\kern-.125emS}}

\title{String effects in SU(2) lattice gauge theory}
\author{M. Caselle\address[TO]{Dipartimento di Fisica
 Teorica dell'Universit\`a di Torino and I.N.F.N.,\\
via P.Giuria 1, I-10125 Torino, Italy}, %
        M. Pepe\address{Institute for Theoretical Physics, University of Bern,\\
        Sidlerstrasse 5, CH-3012 Bern, Switzerland}\thanks{Supported by the Schweizerischer Nationalfond}
        and A.Rago\addressmark[TO]\thanks{Speaker at the conference}
} 
\begin{document}
\begin{abstract}
We discuss the effective string picture for the confining regime
of lattice gauge theories at zero and finite temperature. We present
results of extensive Monte Carlo simulations - performed with the
L\"uscher and Weisz algorithm - for SU(2) Yang-Mills theory in 2+1
dimensions. We also address the issue of "string universality" by
comparing our results with those obtained in other lattice gauge
theories.
\end{abstract}
\maketitle
In the confining phase of a pure gauge theory, the correlation function of two Polyakov loops $P(x)$ at a distance $R$ and at a temperature $T=1/L$:
\eq
G(R)\equiv 
\langle P(x)P^\dagger(x+R) \rangle \equiv {\rm e}^{-F(R,L)},
\label{polya}
\en
\noindent
is expected to follow the so called ``area law''
\eq
F(R,L)\sim \sigma L R + k(L).
\label{area}
\en
Here $\sigma$ denotes the string tension and $k(L)$ is a non-universal
constant  depending only on $L$.
In the rough phase one has to add a correction term (``effective
string corrections'') in order to take 
into account the contribution due to the quantum fluctuations of the flux tube. The simplest way to describe these quantum fluctuations is to 
model the displacement of the flux tube from its
rest position with a $2d$ free bosonic field. This leads to the following result:
\eq
F(R,L)\sim  F_q(R,L)=\sigma L R + k(L)+ F^1_q(R,L)~
\label{a+q}
\en
with
\eq
F_q^1(R,L)=(d-2)\log\left({\eta(\tau)}\right);\hskip0.5cm {-i}\tau={L\over 2R},
\label{bos}
\en
\noindent
where $(d-2)$ 
is the number of transverse dimensions and $\eta(\tau)$ denotes the Dedekind 
eta function
\eq
\eta(\tau)=q^{1\over24}\prod_{n=1}^\infty(1-q^n);\hskip0.5cm q=e^{2\pi i\tau}.
\label{etafun}
\en
The labels $q$ and $1$ in $F^1_q$ recall that this is the first order term of the expansion in $(\sigma L R )^{-1}$ of the flux tube quantum fluctuations.
Eq.(\ref{a+q}) is referred to as the ``free bosonic string approximation''.
At large enough temperatures, higher order terms of  the expansion become important and cannot be neglected. These terms depend on the choice of
the effective string action. One of the simplest proposal \cite{df83,chp03,chpnew} is the Nambu--Goto action, which leads to the following subleading contribution
\eq
F_q^{2}(R,L)=-\frac{\pi^2 L}{1152\ \sigma
R^3}\left[2
E_4(\tau)-E_2^2(\tau)\right],
\label{nlo}
\en
where $E_2$ and $E_4$ are the second and fourth order Eisenstein functions.
 In addition to these corrections a ``boundary term''\cite{lw02} may also be
expected, due to the presence of the Polyakov loops. 
This term gives a contribution equal to the pure free string action provided one replaces 
the interquark distance $R$ with~\cite{chpnew}
\eq
R \to R^*=\frac{R}{(1+2\frac{b}{R})^{\frac12}},
\en
where $b$ is a parameter. 

The observable which better describes the effective string properties is the  
 the combination
\eq
c(R)\equiv -\frac12 R^3\frac1L\log\left(\frac{G(R+1)G(R-1)}{G(R)^2}\right),
\label{defc}
\en
which was recently studied in~\cite{lw02} and \cite{maj}.
\vspace{-1.5mm}\section{The model.}
\vspace{-1mm}\label{sect2}
We studied the pure SU(2) gauge model, with the standard Wilson action:
\eq
S=\beta\sum_p(1-\frac12{\mbox{Tr}~U_p})
\label{wilson}
\en
defined on a 2+1 dimensional cubic lattice of L lattice spacings
in the temporal direction and $N_s$ in the spatial ones. We impose
periodic boundary conditions in the temporal direction in order to
consider the SU(2) Yang-Mills theory at finite temperature. 
The parameters $\beta$ and $L$ are related to the dimensionful gauge
coupling $1/g^2$  and to the temperature $T$ as follows
\eq
\frac{4}{g^2}=a\beta~,~~~~~~ T=\frac{1}{La}.
\en
Here $a$ is the lattice spacing.
\vspace{-1.5mm}\section{Simulation Settings}
\vspace{-1mm}We have performed numerical simulations at $\beta=9$, corresponding to a lattice spacing $a\sim 0.072 fm$ in physical units. 
Exploiting the recent algorithm proposed by L\"uscher and Weisz \cite{Luscher:2001up}, we have
measured with high accuracy the Polyakov loop correlation function
in two different temperature regimes. 

First, we have studied the low 
temperature regime, in which we consider very large values 
of $L$ (we chose to study $L=42,48,54$ and $60$)
 so as to make the finite temperature corrections to the interquark potential
negligible. 
This allows a high precision test both of the free bosonic
string limit (for large values of $R$) 
and of the higher order corrections (for intermediate values of $R$).

Second, we have considered an intermediate temperature regime in which we chose $L=8$ which corresponds to 
$T/T_c\sim 3/4$. 
\vspace{-1.5mm}\section{Analysis of the low temperature data.}
\begin{figure}[!t]
\centering
\includegraphics[width=7.5cm]{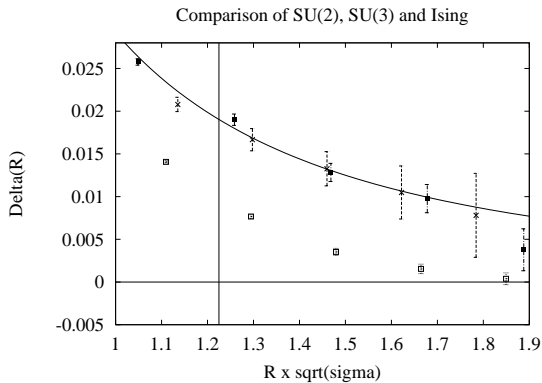}
\vspace{-.8cm}
\caption{Comparison of $c(R)-\pi/24/(1-R^{-2})$ for several 3$d$ gauge theories:
SU(3) Yang-Mills theory (empty squares), the Ising gauge model (filled
squares) and our SU(2) sample with $L=42$ 
(crosses). The continuous line is a fit obtained assuming a $1/R^3$
correction. The straight vertical line denotes $R_c$.}
\label{su2zoom}
\vspace{-.7cm}
\end{figure}

\vspace{-1mm}In fig.\ref{su2zoom} we plot the data obtained for
$c(R)-\frac{\pi}{24(1-R^{-2})}$ as a function of $R\sqrt{\sigma}$. 
The data seem to be well described by a $1/R^3$ behavior;
however a $1/R^2$ correction cannot be completely excluded.

The analysis of the data (see~\cite{cpr} for further details) shows that:\\
-There is an excellent agreement with those obtained for this same model in~\cite{maj},\\-There is a smooth convergence to the expected free bosonic string value as $R$ increases.\\-Higher order corrections are needed to describe our data properly.\\-At intermediate distances $R_c < R < 2/\sqrt{\sigma}$ - $R_c$ being the validity threshold of the effective string picture (see~\cite{chp03}) - the data seem to be well described by a $1/R^3$ type corrections.\\-If a boundary term exists, it is quite small.\\
It is important to stress that while the power of the higher order correction that we find is that predicted by the Nambu-Goto model the numerical coefficient is definitely different from the predicted one. This does not necessarily mean that the Nambu-Goto string is the wrong choice (above all in view of the impressive success in the intermediate temperature regime, see below and~\cite{chp03}). The disagreement could be due to the presence for small values of $R$ of contributions due to irrelevant operators which are not related to the effective string. Further investigations are needed to better understand this point.
\vspace{-1.5mm}\section{Analysis of the intermediate temperature data.}
\begin{figure}[!t]
\centering
\includegraphics[width=7.5cm]{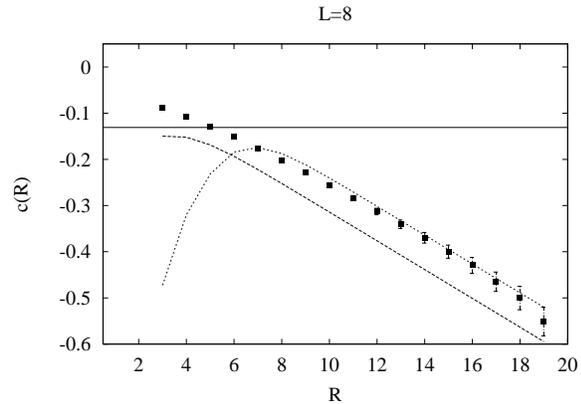}
\vspace{-.8cm}
\caption{Comparison of the results of our simulation for $L=8$ with the
theoretical expectations for the free bosonic string (dashed line) and
for the Nambu-Goto string
truncated at the second order (dotted line). The horizontal continuous
line is the coefficient of the L\"uscher term $\pi/24$.} 
\label{figcl8}
\vspace{-.7cm}
\end{figure}

\vspace{-1mm}In this section we report on the results we have obtained in the
intermediate temperature regime. In fig.\ref{figcl8} we plot our data
for $c(R)$ together with the theoretical expectations of the pure
bosonic and of the Nambu-Goto string. 

The analysis of the data shows that:

-The L\"uscher term alone $c(R)=\pi/24$ does not properly describe the behavior
of the data. The whole functional form of the correction (see eq.(\ref{bos}) and the dashed line in fig.\ref{figcl8}) is needed.

-The figure shows an impressive agreement between our data and the truncated Nambu-Goto prediction (see eq.(\ref{nlo})). Interestingly, a similar
remarkable agreement has also been recently observed in the $3d$ Ising model
in the same intermediate temperature regime.

\vspace{-1.5mm}\section{Comparison with SU(3) and Ising gauge models.}
\vspace{-1mm}\label{sect6}
It is very interesting to compare our results with those recently obtained for 
the SU(3)~\cite{lw02} and the Ising \cite{chpnew} gauge models in 3 dimensions.

In fig.\ref{su2zoom}, we plot the three data sets as a function of the dimensionless
quantity $R\sqrt{\sigma}$. We can see that the SU(2) and the Ising data are
quite close and they show rather strong deviations from the pure free string behaviour. 
 This might suggest that they are described by the same
 effective string correction even beyond  the free non-interacting limit. 

On the contrary the SU(3) data show a rather different behaviour, with a 
much smaller deviation from the free
string behaviour. This may suggest that the $3d$ SU(3) Yang-Mills theory belongs to a different string universality  
class. It can be interesting to study Yang-Mills theories with other
gauge groups in order to see if there are common patterns or trends w.r.t. the
effective string description. The natural choice in this respect is
to consider the group Sp(2). We plan to address this issue in a
future study.

Finally, we would like to emphasize that, although the effective string description
represents very successfully some aspects of the interquark potential,
some other issues - for instance the spectrum of the excited states
\cite{Jug03} - need to be understood.
\vspace{-1.5mm}\section*{Acknoledgements}
\vspace{-1mm}We acknowledge useful discussions with F.Gliozzi, J.Juge, J.Kuti, C.Morningstar and U.-J. Wiese.


\begin{thebibliography}{99}

\bibitem{df83}
K.~Dietz and T.~Filk,
Phys.\ Rev.\ D {\bf 27} (1983) 2944.

\bibitem{chp03}
M.~Caselle, M.~Panero and P.~Provero,
JHEP {\bf 0206} (2002) 061
[arXiv:hep-lat/0205008].

M.~Caselle, M.~Hasenbusch and M.~Panero,
JHEP {\bf 0301} (2003) 057
[arXiv:hep-lat/0211012].

\bibitem{chpnew}
M.~Caselle, M.~Hasenbusch and M.~Panero,
this proceeding.

\bibitem{lw02}
M.~L\"uscher and P.~Weisz,
JHEP {\bf 0207} (2002) 049
[arXiv:hep-lat/0207003].

\bibitem{maj}
P.~Majumdar,
arXiv:hep-lat/0211038.

\bibitem{Luscher:2001up}
M.~L\"uscher and P.~Weisz,
JHEP {\bf 0109}, 010 (2001)
[arXiv:hep-lat/0108014].

\bibitem{cpr}
M.~Caselle, M.~Pepe and A. Rago,
in preparation.

\bibitem{Jug03}
K.~J.~Juge, J.~Kuti and C.~Morningstar,
Phys.\ Rev.\ Lett.\  {\bf 90} (2003) 161601.
\end{thebibliography}
\end{document}